\documentclass[preprint,authoryear,12pt]{elsarticle}
\usepackage{amsmath}
\usepackage{amsmath,color}
\usepackage{graphicx}

\begin{document}

\begin{frontmatter}


\title{A Robust t-process  Regression Model with Independent Errors}

\author[a,b]{Zhanfeng Wang}
\address
[a]{Department of Statistics and Finance, The School of Management, University of Science and Technology of China, Hefei, China}

\author[c]{Maengseok Noh}
\address
[c]{Department of Statistics, Pukyong National University, Busan, Korea}

\author[b]{Youngjo Lee\corref{cor1}}
\cortext[cor1]{Corresponding author. Email: youngjo@snu.ac.kr}
\address[b]{Department of Statistics, Seoul National University, Seoul, Korea}

\author[d]{Jian Qing Shi}
\address
[d]{School of Mathematics and Statistics, Newcastle University, Newcastle, UK}

\begin{abstract}
Gaussian process regression (GPR) model is well-known to be susceptible to outliers.
Robust process regression models based on t-process or other heavy-tailed processes have been developed to address the problem. However, due to the nature of the current definition for heavy-tailed processes, the unknown process regression function and the random errors are always defined jointly and thus dependently. This definition, mainly owing to the dependence assumption involved, is not justified in many practical problems and thus  limits the application of those robust approaches. It also results in a limitation of the theory of robust analysis. In this paper, we propose a new robust process regression model enabling independent random errors. An efficient estimation procedure is developed.
Statistical properties, such as unbiasness  and information consistency,  are provided. Numerical studies show that the proposed method is robust against outliers and has a better performance in prediction compared with the existing models. We illustrate that the estimated random-effects are useful in detecting outlying curves.
\end{abstract}
\end{frontmatter}

\section{Introduction}

In regression analysis, we are interested in modelling the relationship
between response $y$ and covariate $\mbox{\boldmath ${x}$}\in \mathcal{X}%
\subset R^{p}$. A nonparametric regression usually uses $E(y|%
\mbox{\boldmath
${x}$})$ to fit $y$, based on the model $y=E(y|\mbox{\boldmath ${x}$})+e,$
where $e$ is an error term. Let $f_{0}(\mbox{\boldmath ${x}$})=E(y|%
\mbox{\boldmath ${x}$})$ be a fixed unknown function of $%
\mbox{\boldmath
${x}$}$. Then, the nonparametric regression model is rewritten as
\begin{equation}
{\label{model2-1}}y=f_{0}(\mbox{\boldmath ${x}$})+e.
\end{equation}
To estimate function $f_{0}$, this paper considers a process regression
model
\begin{equation}
y(\mbox{\boldmath ${x}$})=f(\mbox{\boldmath${x}$})+\epsilon (%
\mbox{\boldmath
${x}$}),  \label{model2-2}
\end{equation}
where $f(\mbox{\boldmath${x}$})$ is a random function and $\epsilon (%
\mbox{\boldmath
${x}$})$ is an error process. In model (\ref{model2-2}), $f$ can be treated
as a nonparametric random effect and thus this model can be called a
nonparametric random effect functional regression model.

When $f$ and $\epsilon $ are independent Gaussian processes (GPs), the model
(\ref{model2-2}) is called Gaussian process regression (GPR) model. The
details about the GPR model can be found in \cite{r7}, \cite{r8}. Recent
developments include GPR analysis for batch data \citep{r9}, GPR for
single-index model \citep{r1} and generalized GPR for non-Gaussian
functional data \citep{r10}. However, it is well-known that the GPR model is
susceptible to outliers. To overcome this problem, robust methods are
developed based on t-process and other heavy tailed processes; for example,
\cite{Shah14} used a simple t-process to replace a GP; \cite{r11} proposed
an extended t-process regression model (eTPR); and \cite{Cao17} developed
robust models based on other heavy-tailed processes such as Slash process
and contaminated-normal process. Heavy-tailed processes, particularly
t-process, have been used frequently in many different areas to build a
robust model, for example, \cite{r13} and \cite{r14} used t-process to build
a multi-task learning model, and \cite{r12} employed matrix-variate
t-process and a variational approximation method to construct a sparse
matrix-variate block model. However, in these robust process regression
models, the unknown regression function $f$ is defined jointly with random
errors $\epsilon $, and thus they are dependent. %
Although this brings technical convenience in implementation, the dependence
assumption may not be justified in many practical problems and is also not
necessary in developing a theory.

In model \eqref{model2-2}, the regression function $f$ is the main part,
describing the regression relationship between the response variable $y$ and
the covariates $\mbox{\boldmath ${x}$}$. The estimation of unknown $f$
converges to the true function $f_{0}(\mbox{\boldmath ${x}$})=E(y|%
\mbox{\boldmath ${x}$})$ when a GPR model is assumed as the sample size
tends to infinity \citep[see e.g.][]{Choi07,r15,r8}. The error $\epsilon (%
\mbox{\boldmath ${x}$})$ is usually not dependent on the covariates,
otherwise the dependent part can be merged with $f(\mbox{\boldmath ${x}$})$.
Purely because of technical convenience, $f$ and $\epsilon $ are defined
jointly when heavy-tailed processes are used for robust approaches. In this
paper, we propose a new approach in which $f$ and $\epsilon $ are separately
modeled using extended t-processes (ETPs), not jointly. Hereafter, we name
this new robust model as \emph{independent error model} while the existing
models as \emph{joint error models}.

When $f\sim GP$ and $\epsilon\sim GP$, the joint error model is the same as
the independent error model (GPR model) because the sum of two GPs is again
a GP. However, in general, the independent error model behaves differently
from the joint error model. The former is more flexible and suited to
practical application as we shall show.
We will also show that the independent error model with TP errors is more
robust than the corresponding joint error models, and the function estimator
is less sensitive to outlying curves.

The independent error models however involve
intractable integrations in the calculation of predictive mean and variance.
To address the problem, an efficient estimation procedure is developed via
h-likelihood \citep[see e.g.][]{r6}. We also extend the idea to build a
process regression model for batched functional data. Statistical
properties, such as unbiasness and information consistency, will be shown.
Simulation studies show that the proposed method is robust against outlying
curves, and application to real data demonstrates that the proposed method
provides stable results no matter whether data consist of observations from
subject with odd responses. In the research, we also have an interesting
finding: the values of estimated random effects can be used to detect
outlying curves.

The remainder of the paper is organized as follows. Section 2 presents an
independent error regression model and studies predictive distribution of
function $f$. In section 3, a general functional regression model for batch
data is studied and an h-likelihood estimation procedure is proposed to
calculate the prediction. Numerical studies and real examples, including
detection of outlying curves, are given in Section 4. All the proofs are
presented in Appendix.

\section{Independent error regression models}

To study a robust independent error regression model, we introduce an
extended t-process (ETP) as follows. For a random function $f(\cdot ):%
\mathcal{X}\rightarrow R$, if
\begin{equation*}
f|r\sim GP(h,rk)~~{\mbox{and}}~~r\sim IG(\nu ,\nu -1),
\end{equation*}
then $f$ follows an ETP, denoted by $f\sim ETP(\nu ,h,k)$, where $h(\cdot ):%
\mathcal{X}\rightarrow R$ is a mean function, $k(\cdot ,\cdot ):\mathcal{X}%
\times \mathcal{X}\rightarrow R$ is a covariance kernel, $GP(h,k)$ stands
for a Gaussian process with mean function $h$ and covariance kernel $k$, and
$IG(\nu ,\nu -1)$ is an inverse gamma distribution with parameter $\nu $ and
the density function of
\begin{equation*}
g_{\nu }(r)=\frac{1}{\Gamma (\nu )}(\frac{\nu -1}{r})^{\nu +1}\frac{1}{\nu -1%
}\exp {(-\frac{\nu -1}{r})},
\end{equation*}
having $E(r)=1$ and $Var(r)=1/(\nu -2)$. Here, $r$ is a random effect,
affecting the covariance kernel of the GP.

In functional regression, it is convenient to set $h=0$. Thus, we set
\begin{equation*}
GP(0,k)=GP(k)\text{ and }ETP(\nu ,0,k)=ETP(\nu ,k)
\end{equation*}
if no confusion occurs. Note that\ $GP(k)=ETP(\infty ,k)$\ with a constant $%
r=1.$ In model (\ref{model2-2}), suppose that $f$ and $\epsilon $ are
independent random processes, for example,

\begin{itemize}
\item  GP-GP model: $f\sim GP(k)$ and $\epsilon \sim GP(k_{\epsilon })$;

\item  GP-TP model: $f\sim GP(k)$ and $\epsilon \sim ETP(\nu
_{1},k_{\epsilon })$;

\item  TP-TP model: $f\sim ETP(\nu _{0},k)$ and $\epsilon \sim ETP(\nu
_{1},k_{\epsilon })$;

\item  TP-GP model: $f\sim ETP(\nu _{0},k)$ and $\epsilon \sim
GP(k_{\epsilon })$,

\end{itemize}

where $k$ is a covariance kernel for the process $f$ and $k_{\epsilon}$ is
that for the process $\epsilon$. We usually set $k_{\epsilon}(%
\mbox{\boldmath ${u}$},\mbox{\boldmath ${v}$})=\phi I(\mbox{\boldmath ${u}$}=%
\mbox{\boldmath
${v}$})$ for $\mbox{\boldmath ${u}$}$, $\mbox{\boldmath ${v}$}\in\mathcal{X}$%
. Here $I(\cdot )$ is an indicator function.

Let ${\mathcal{D}}=\{(y_k,\mbox{{\boldmath ${x}$}$_{k}$}),k=1,...,n\}$ be
the observed data set from model (\ref{model2-1}), where $y_k=y(%
\mbox{{\boldmath
${x}$}$_{k}$})$. The GP-GP model is the well-known GPR model with an
explicit conditional prediction process of $f|{\mathcal{D}}$. Note that in
the GPR model $f+\epsilon \sim GP(k+k_{\epsilon})$. However, the conditional
prediction process of $f|{\mathcal{D}}$ does not have close forms for the
rest of models, and it actually involves intractable integrations. We will
propose an efficient implementation method via h-likelihood and will use the
TP-TP model as an example to illustrate the idea.

\subsection{Predictive distributions}

Based on the construction of ETP, suppose that $f$ and $\epsilon$ are
generated by,
\begin{align}
& f|r_{0}\sim GP(r_{0} k),~~r_{0}\sim \mathrm{IG}(\nu_0,\nu_0-1),  \notag \\
& \epsilon|r_{1}\sim GP(r_{1} k_{\epsilon}),~~r_{1}\sim \mathrm{IG}%
(\nu_1,\nu_1-1),  \notag \\
& r_{0} \text{ and } r_{1} \text{ are independent.}  \notag
\end{align}
When $r_0 = r_1$, we have
\begin{equation*}
f+\epsilon | r_0 \sim \mathrm{GP}(r_0 k + r_0 k_{\epsilon}),\text{ } r_0
\sim \mathrm{IG} (\nu_0,\nu_0-1).
\end{equation*}
Then, the resulting process for $f+\epsilon$ is the eTPR, a joint error
model. Thus, conditional on $r_{0}$, the sum of the eTPRs becomes an eTPR
again. This setting is convenient for implementation and makes the
derivation of the theory easy, but has a drawback as we shall show.

In the TP-TP model,
\begin{align}
& y(\cdot)|f,r_1\sim GP\left( f,~r_1 k_{\epsilon}\right),  \notag \\
& y(\cdot) |r_0,r_1 \sim GP\left(0,~r_{0} k+r_1
k_{\epsilon}\right)=GP\left(r_{0} k+r_1 k_{\epsilon}\right),  \notag \\
& r_{0}\sim \mathrm{IG}(\nu_0,\nu_0-1),~r_{1}\sim \mathrm{IG}(\nu_1,\nu_1-1),
\notag
\end{align}
where $y(\cdot)$ is the response function. Parameters $\nu_0$ and $\nu_1$
are pre-specified here, but they can be estimated in the model for batch
data as we shall show.

When ${\mbox{\boldmath ${r}$}}=(r_{0},r_{1})^{T}$ is given, $y$ and $f$ have
a GPR model. Thus, the results for the GPR model can be extended to the
above model given $\mbox{\boldmath ${r}$}$. For the observed data ${\mathcal{%
D}}$, we have
\begin{align}
& \mbox{{\boldmath ${f}$}}(\mbox{{\boldmath ${X}$}})|r_{0},%
\mbox{{\boldmath
${X}$}}\sim N(0,~r_{0}\mbox{{\boldmath ${K}$}$_{n}$}),  \label{fndist-1} \\
& \mbox{{\boldmath ${y}$}}|\mbox{{\boldmath ${f}$}},\mbox{{\boldmath ${r}$}},%
\mbox{{\boldmath ${X}$}}\sim N\left( \mbox{{\boldmath ${f}$}$_{n}$},~\phi
r_{1}\mbox{{\boldmath
${I}$}$_{n}$}\right) ,  \label{yncdist-1} \\
& \mbox{{\boldmath ${y}$}}|\mbox{{\boldmath ${r}$}},\mbox{{\boldmath ${X}$}}%
\sim N(0,~\mbox{{\boldmath ${C}$}$_{r}$}),  \label{yndist-1}
\end{align}
where $\mbox{\boldmath ${X}$}=(\mbox{{\boldmath ${x}$}$_{1}^{T}$},...,%
\mbox{{\boldmath ${x}$}$_{n}^{T}$})^{T}$, $\mbox{\boldmath ${y}$}%
=(y_{1},...,y_{n})^{T}$, $\mbox{{\boldmath ${f}$}$_{n}$}=%
\mbox{{\boldmath
${f}$}}(\mbox{{\boldmath ${X}$}})=(f(\mbox{{\boldmath ${x}$}$_{1}$}),...,f(%
\mbox{{\boldmath ${x}$}$_{n}$}))^{T}$, $\mbox{{\boldmath
${K}$}$_{n}$}=(k_{kl})_{n\times n}$ with $k_{kl}=k(%
\mbox{{\boldmath
${x}$}$_{k}$},\mbox{{\boldmath${x}$}$_{l}$})$, and $%
\mbox{{\boldmath
${C}$}$_{r}$}=r_{0}\mbox{{\boldmath
${K}$}$_{n}$}+\phi r_{1}\mbox{{\boldmath ${I}$}$_{n}$}$. From (\ref{fndist-1}%
)-(\ref{yndist-1}), we have
\begin{equation*}
f(\mbox{{\boldmath ${X}$}})|\mbox{\boldmath ${r}$},{\mathcal{D}}\sim N(%
\mbox{{\boldmath ${\mu}$}$_{r}$},\mbox{{\boldmath
${\Sigma}$}$_{r}$})
\end{equation*}
with
\begin{align}
& \mbox{{\boldmath ${\mu}$}$_{r}$}=E(f(\mbox{{\boldmath ${X}$}})|%
\mbox{\boldmath ${r}$},{\mathcal{D}})=r_{0}\mbox{{\boldmath ${K}$}$_{n}$}%
\mbox{{\boldmath
${C}$}$_{r}^{-1}$}\mbox{{\boldmath ${y}$}},  \label{cond.pred-1} \\
& \mbox{{\boldmath ${\Sigma}$}$_{r}$}=Var(f(\mbox{{\boldmath ${X}$}})|%
\mbox{\boldmath ${r}$},{\mathcal{D}})=r_{0}\mbox{{\boldmath
${K}$}$_{n}$}-r_{0}^{2}\mbox{{\boldmath ${K}$}$_{n}$}%
\mbox{{\boldmath
${C}$}$_{r}^{-1}$}\mbox{{\boldmath ${K}$}$_{n}$}.  \label{cond.cov-1}
\end{align}

For a new data point $\mbox{\boldmath ${z}$}$, let $%
\mbox{{\boldmath
${k}$}$_{z}$}=(k(\mbox{\boldmath ${z}$},\mbox{{\boldmath ${x}$}$_{1}$}%
),...,k(\mbox{\boldmath ${z}$},\mbox{{\boldmath ${x}$}$_{n}$}))^T$. Similar
to (\ref{cond.pred-1}) and (\ref{cond.cov-1}) it is easy to show
\begin{equation*}
f(\mbox{{\boldmath ${z}$}})|\mbox{\boldmath ${r}$}, {\mathcal{D}}\sim N(%
\mbox{{\boldmath ${\mu}$}$_{r}^{*}$},\mbox{{\boldmath ${\Sigma}$}$_{r}^{*}$})
\end{equation*}
with
\begin{align}
&\mbox{{\boldmath ${\mu}$}$_{r}^{*}$}=E(f(\mbox{{\boldmath ${z}$}})|%
\mbox{\boldmath ${r}$},{\mathcal{D}})=r_{0}
\mbox{{\boldmath
${k}$}$_{z}^{T}$} \mbox{{\boldmath
${C}$}$_{r}^{-1}$}\mbox{{\boldmath ${y}$}},  \notag \\
&\mbox{{\boldmath ${\Sigma}$}$_{r}^{*}$}=Var(f(\mbox{{\boldmath ${z}$}} )|%
\mbox{\boldmath ${r}$},{\mathcal{D}})=r_{0}k(\mbox{\boldmath ${z}$},%
\mbox{\boldmath ${z}$})-r_{0}^2\mbox{{\boldmath ${k}$}$_{z}^{T}$}%
\mbox{{\boldmath ${C}$}$_{r}^{-1}$}\mbox{{\boldmath ${k}$}$_{z}$}.  \notag
\end{align}

\vskip10pt \noindent \textbf{Remark 1} Let $\mbox{{\boldmath ${C}$}}=%
\mbox{{\boldmath ${K}$}$_{n}$}+\phi \mbox{{\boldmath ${I}$}$_{n}$}$. When $%
r_{0}=r_{1}=r$, the model becomes an eTPR joint error model in \cite{r11}.
From (\ref{cond.pred-1}) we have
\begin{equation*}
E(f(\mbox{{\boldmath ${X}$}})|\mbox{\boldmath ${r}$},\mathcal{D})=%
\mbox{{\boldmath
${K}$}$_{n}$}\mbox{{\boldmath ${C}$}$^{-1}$}\mbox{{\boldmath
${y}$}},
\end{equation*}
which does not depend on random effect $\mbox{\boldmath ${r}$}$ and is
exactly the conditional mean of $f(\mbox{{\boldmath ${X}$}})|\mathcal{D}$
from a GPR model. From (\ref{cond.pred-1}) and (\ref{cond.cov-1}) we have
\begin{align}
Var(f(\mbox{{\boldmath ${X}$}})|\mathcal{D})=& E\{Var(f(%
\mbox{{\boldmath
${X}$}})|\mbox{\boldmath ${r}$},\mathcal{D})|\mathcal{D}\}  \notag \\
=& s_{0}\left( \mbox{{\boldmath ${K}$}$_{n}$}-\mbox{{\boldmath ${K}$}$_{n}$}%
\mbox{{\boldmath${C}$}$^{-1}$}\mbox{{\boldmath
${K}$}$_{n}$}\right) ,  \notag
\end{align}
where
\begin{equation*}
s_{0}=E(r|\mathcal{D})=\frac{2(\nu _{0}-1)+\mbox{{\boldmath ${y}$}$^{T}$}%
\mbox{{\boldmath ${C}$}$^{-1}$}\mbox{{\boldmath ${y}$}}}{2(\nu _{0}-1)+n}.
\end{equation*}
$s_{0}$ stands for the difference in predictive variance between the GPR
model and the eTPR joint error model.
As $n\rightarrow \infty $, $s_{0}\rightarrow 1$, consequently the eTPR joint
error model tends to the GPR. Thus, the robustness of the joint error model
is deteriorated when $n$ is large. By contrast, 
in the independent error models with $r_{0}\neq r_{1}$, both mean and
variance are different from those in the joint models; see (\ref{cond.pred-1}%
) and (\ref{cond.cov-1}), where mean and variance depend on $r_{0}$ and $%
r_{1}$. This makes the independent error model to be robust even when $n$ is
large, resulting in a better performance than the joint error models in the
presence of outliers.

\section{General independent error models for batch data}

More generally, model (\ref{model2-1}) can be extended to a functional
regression model for batch functional data,
\begin{equation}
y_{ijk}=f_{0i}(\mbox{\boldmath${x}$}_{ijk})+\epsilon
_{ijk},~~i=1,...,I,j=1,...,J,k=1,...n,  \label{true}
\end{equation}
where $y_{ijk}$ is the $k$th observed data under the $j$th curve in the $i$%
the group, $f_{0i}(\mbox{\boldmath${x}$}_{ijk})$ is the value of unknown
function $f_{0i}(\cdot )$ at the $p\times 1$ observed covariate $%
\mbox{{\boldmath
${x}$}$_{ijk}$}\in \mathcal{X}\subset R^{p}$ and $\epsilon _{ijk}$ is an
error term. In the old (young) dataset which is discussed in Section 4,
there are $I=2$ groups, $J=12$ or 13 subjects and $n=180$ observed times.

To estimate true unknown functions $f_{0i}$, we consider a process
regression model
\begin{equation}
y_{ij}(\mbox{\boldmath ${x}$})=f_{i}(\mbox{\boldmath${x}$})+\epsilon _{ij}(%
\mbox{\boldmath
${x}$}),~~i=1,...,I,j=1,...,J,  \label{assumed}
\end{equation}
where $f_{i}(\mbox{\boldmath${x}$})$ is a random function, $\epsilon _{ij}(%
\mbox{\boldmath
${x}$})$ is an error process for $\mbox{\boldmath ${x}$}\in \mathcal{X}$, $%
y_{ijk}=y_{ij}(\mbox{{\boldmath ${x}$}$_{ijk}$})$ and $\epsilon
_{ijk}=\epsilon _{ij}(\mbox{{\boldmath ${x}$}$_{ijk}$})$. In the TP-TP
model, we assume $f_i$ and $\epsilon_{ij}$ are independent and
\begin{equation*}
f_{i}\sim ETP(\nu _{0},k_{i}) {\mbox{~and~}} \epsilon _{ij}\sim ETP(\nu
_{1},k_{\epsilon i}),~~i=1,...,I
\end{equation*}
where $k_i$ is a covariance kernel and $k_{\epsilon i}(\mbox{\boldmath ${u}$}%
,\mbox{\boldmath ${v}$})=\phi_i I(\mbox{\boldmath ${u}$}=%
\mbox{\boldmath
${v}$})$ for $\mbox{\boldmath ${u}$}$, $\mbox{\boldmath ${v}$}\in\mathcal{X}$%
. Under this setup with $J=1$, \cite{r11} discussed a joint error eTPR
model. Similar to the discussion in Section 2, $f_i$ and $\epsilon_{ij}$ can
be defined by,
\begin{align}
& f_i|r_{i0}\sim GP(r_{i0} k_i),~~r_{i0}\sim \mathrm{IG}(\nu_0,\nu_0-1),
\notag \\
& \epsilon_{ij}|r_{ij}\sim GP(r_{ij} k_{\epsilon i}),~~r_{ij}\sim \mathrm{IG}%
(\nu_1,\nu_1-1),  \notag \\
& r_{i0}, r_{i1},...,r_{iJ} \text{ are independent,} ~~i=1,...,I.  \notag
\end{align}
When $I=J=1$, parameters $\nu_0$ and $\nu_1$ are not estimable, because
there are only two random effects $r_{10}$ and $r_{11}$ such that $r_{10}$
and $r_{11}$ are confounded with the covariance kernels $k_1 $ and $%
k_{\epsilon 1}$. Following \cite{r11}, when $I=J=1$, a convenient way is to
set $\nu_0=\nu_1=1.05$. 
When $I>1$, $\nu_0$ and $\nu_1 $ are estimable.

Without loss of generality, we set $\mbox{{\boldmath ${x}$}$_{i1k}$}=...=%
\mbox{{\boldmath ${x}$}$_{iJk}$}=\mbox{{\boldmath ${x}$}$_{ik}$}$ $%
(i=1,...,I,k=1,...,n)$, which means the same observed covariates $\{%
\mbox{{\boldmath ${x}$}$_{i1}$},...,\mbox{{\boldmath ${x}$}$_{in}$}\}$ for $%
J $ different curves in the $i$th group. Let ${\mbox{\boldmath ${X}$}}_{i}=(%
\mbox{{\boldmath ${x}$}$_{i1}$},...,\mbox{{\boldmath ${x}$}$_{in}$})^{T}$, $%
\mbox{{\boldmath ${y}$}$_{ij}$}=(y_{ij}(\mbox{{\boldmath ${x}$}$_{i1}$}%
),...,y_{ij}(\mbox{{\boldmath ${x}$}$_{in}$}))^{T}$, ${\mbox{\boldmath ${y}$}%
_{i}}=(\mbox{{\boldmath ${y}$}$_{i1}^{T}$},...,%
\mbox{{\boldmath
${y}$}$_{iJ}^{T}$})^{T}$ and ${\mathcal{D}}_{n}=\{%
\mbox{{\boldmath
${X}$}$_{i}$},\mbox{{\boldmath ${y}$}$_{i}$},i=1,...,I\}$.

\subsection{Predictive distributions}

For model (\ref{assumed}), we have
\begin{align}
& \mbox{{\boldmath ${Y}$}$_{i}$}|f_{i},\mbox{{\boldmath ${r}$}$_{i}$}\sim
GP\left( \mbox{{\boldmath ${b}$}$_{J}$}\bigotimes f_{i},~%
\mbox{{\boldmath
${r}$}$_{ei}$}\bigotimes k_{\epsilon i}\right) ,  \notag \\
& \mbox{{\boldmath ${Y}$}$_{i}$}|\mbox{{\boldmath ${r}$}$_{i}$}\sim GP\left(
0,~r_{i0}\mbox{\boldmath ${A}$}\bigotimes k_{i}+%
\mbox{{\boldmath
${r}$}$_{ei}$}\bigotimes k_{\epsilon i}\right) =GP\left( r_{i0}%
\mbox{\boldmath ${A}$}\bigotimes k_{i}+\mbox{{\boldmath
${r}$}$_{ei}$}\bigotimes k_{\epsilon i}\right) ,  \notag \\
& r_{i0}\sim \mathrm{IG}(\nu _{0},\nu _{0}-1),~r_{ij}\sim \mathrm{IG}(\nu
_{1},\nu _{1}-1),~j=1,...,J,  \notag
\end{align}
where $\mbox{{\boldmath ${Y}$}$_{i}$}=(y_{i1}(\cdot ),...,y_{iJ}(\cdot ))^{T}
$, $\mbox{{\boldmath ${r}$}$_{i}$}=(r_{i0},r_{i1},...,r_{iJ})^{T}$, $%
\mbox{{\boldmath ${r}$}$_{ei}$}$ is a $J\times J$ diagonal matrix with
diagonal components $\{r_{i1},...,r_{iJ}\}$, $\mbox{{\boldmath ${b}$}$_{J}$}%
=(1,...,1)^{T}$ is a $J$-length vector of 1's, $\mbox{\boldmath
${A}$}$ is a $J\times J$ matrix with all elements of 1, and $\bigotimes $
stands for Kronecker product. Denote that ${\mbox{\boldmath ${r}$}}=(%
\mbox{{\boldmath ${r}$}$_{1}^{T}$},...,\mbox{{\boldmath ${r}$}$_{I}^{T}$}%
)^{T}$. Based on the observed data ${\mathcal{D}}_{n}$, we can show that
\begin{align}
& \mbox{{\boldmath ${f}$}$_{in}$}=\mbox{{\boldmath ${f}$}$_{i}$}(%
\mbox{{\boldmath ${X}$}$_{i}$})|r_{i0},\mbox{{\boldmath ${X}$}$_{i}$}\sim
N(0,~r_{i0}\mbox{{\boldmath ${K}$}$_{in}$}),  \notag \\
& \mbox{{\boldmath ${y}$}$_{i}$}|\mbox{{\boldmath ${f}$}$_{i}$},%
\mbox{{\boldmath ${r}$}$_{i}$},\mbox{{\boldmath ${X}$}$_{i}$}\sim N\left( %
\mbox{{\boldmath ${b}$}$_{J}$}\bigotimes \mbox{{\boldmath ${f}$}$_{in}$}%
,~\phi _{i}\mbox{{\boldmath ${r}$}$_{ei}$}\bigotimes
\mbox{{\boldmath
${I}$}$_{n}$}\right) ,  \notag \\
& \mbox{{\boldmath ${y}$}$_{i}$}|\mbox{{\boldmath ${r}$}$_{i}$},%
\mbox{{\boldmath ${X}$}$_{i}$}\sim N(0,~\mbox{{\boldmath ${C}$}$_{ri}$}),
\notag
\end{align}
where $\mbox{{\boldmath ${f}$}$_{i}$}(\mbox{{\boldmath ${X}$}$_{i}$})=(f_{i}(%
\mbox{{\boldmath ${x}$}$_{i1}$}),...,f_{i}(\mbox{{\boldmath ${x}$}$_{in}$}%
))^{T}$, $\mbox{{\boldmath
${K}$}$_{in}$}=(k_{ijl})_{n\times n}$ with $k_{ijl}=k_{i}(%
\mbox{{\boldmath
${x}$}$_{ij}$},\mbox{{\boldmath${x}$}$_{il}$})$, and $%
\mbox{{\boldmath
${C}$}$_{ri}$}=r_{i0}\mbox{\boldmath ${A}$}\bigotimes
\mbox{{\boldmath
${K}$}$_{in}$}+\phi _{i}\mbox{{\boldmath ${r}$}$_{ei}$}\bigotimes %
\mbox{{\boldmath ${I}$}$_{n}$}$.

Again, it gives
\begin{equation*}
f_i(\mbox{{\boldmath ${X}$}$_{i}$})|\mbox{\boldmath ${r}$}, {\mathcal{D}}%
_{n}\sim N(\mbox{{\boldmath ${\mu}$}$_{ri}$},%
\mbox{{\boldmath
${\Sigma}$}$_{ri}$})
\end{equation*}
with
\begin{align}
&\mbox{{\boldmath ${\mu}$}$_{ri}$}=E(f_i(\mbox{{\boldmath ${X}$}$_{i}$})|%
\mbox{\boldmath ${r}$},{\mathcal{D}_n})=r_{i0}(%
\mbox{{\boldmath
${b}$}$_{J}^{T}$}\bigotimes \mbox{{\boldmath ${K}$}$_{in}$})%
\mbox{{\boldmath
${C}$}$_{ri}^{-1}$}\mbox{{\boldmath ${y}$}$_{i}$},  \label{cond.pred} \\
&\mbox{{\boldmath ${\Sigma}$}$_{ri}$}=Var(f_i(\mbox{{\boldmath ${X}$}$_{i}$}%
)|\mbox{\boldmath ${r}$},{\mathcal{D}_n})=r_{i0}
\mbox{{\boldmath
${K}$}$_{in}$}-r_{i0}^2(\mbox{{\boldmath ${b}$}$_{J}^{T}$}\bigotimes %
\mbox{{\boldmath ${K}$}$_{in}$})\mbox{{\boldmath ${C}$}$_{ri}^{-1}$}(%
\mbox{{\boldmath ${b}$}$_{J}$}\bigotimes \mbox{{\boldmath ${K}$}$_{in}$}).
\label{cond.cov}
\end{align}

\vskip10pt \noindent \textbf{Remark 2} Under a special case of $%
r_{i0}=r_{i1}=...=r_{iJ}=r$ which is actually a joint error model, $%
\mbox{{\boldmath ${\mu}$}$_{ri}$}$ in (\ref{cond.pred}) is independent of $%
\mbox{\boldmath ${r}$}$, thus it becomes the conditional mean of $f_{i}(%
\mbox{{\boldmath ${X}$}$_{i}$})|\mathcal{D}_{n}$, the same mean as the one
from a GPR model. Equation (\ref{cond.cov}) shows that
\begin{equation*}
Var(f_{i}(\mbox{{\boldmath ${X}$}$_{i}$})|\mathcal{D}_{n})=s_{0i}\left( %
\mbox{{\boldmath ${K}$}$_{in}$}-(\mbox{{\boldmath
${b}$}$_{J}^{T}$}\bigotimes \mbox{{\boldmath ${K}$}$_{in}$})%
\mbox{{\boldmath
${C}$}$_{i}^{-1}$}(\mbox{{\boldmath ${b}$}$_{m}$}\bigotimes
\mbox{{\boldmath
${K}$}$_{in}$})\right) ,
\end{equation*}
where $\mbox{{\boldmath ${C}$}$_{i}$}=\mbox{{\boldmath ${K}$}$_{in}$}+\phi
_{i}\mbox{{\boldmath ${I}$}$_{n}$}$, and
\begin{equation*}
s_{0i}=E(r|\mathcal{D}_{n})=\frac{2(\nu _{0}-1)+%
\mbox{{\boldmath
${y}$}$_{i}^{T}$}\mbox{{\boldmath ${C}$}$_{i}^{-1}$}%
\mbox{{\boldmath
${y}$}$_{i}$}}{2(\nu _{0}-1)+nJ}.
\end{equation*}
This special case with $J=1$ is the eTPR model discussed in \cite{r11}.

\subsection{Estimation procedure}

In independent error models, $E(f_i(\mbox{\boldmath ${X}$})|%
\mbox{\boldmath
${r}$},\mathcal{D}_n)$ and $Var(f_i(\mbox{\boldmath ${X}$})|%
\mbox{\boldmath
${r}$},\mathcal{D}_n)$ depend on unknown random effect $%
\mbox{\boldmath
${r}$}$. One method to calculate them is to integrate out $%
\mbox{\boldmath
${r}$}$ via conditional distribution of $\mbox{\boldmath ${r}$}|{\mathcal{D}%
_n}$, that is
\begin{align}
&E(f_i(\mbox{\boldmath ${X}$})|{\mathcal{D}_n})=\int E(f_i(%
\mbox{\boldmath
${X}$})|\mbox{\boldmath ${r}$}, {\mathcal{D}_n}) g(\mbox{\boldmath ${r}$}|{%
\mathcal{D}_n}) d\mbox{\boldmath ${r}$}  \notag \\
&Var(f_i(\mbox{\boldmath ${X}$})|{\mathcal{D}_n})=\int \left(Var(f_i(%
\mbox{\boldmath ${X}$})|\mbox{\boldmath ${r}$}, {\mathcal{D}_n})+(E(f_i(%
\mbox{\boldmath ${X}$})|\mbox{\boldmath ${r}$}, {\mathcal{D}_n}))^2\right) g(%
\mbox{\boldmath ${r}$}|{\mathcal{D}_n}) d\mbox{\boldmath
${r}$}  \notag \\
&\hskip 3cm - (E(f_i(\mbox{\boldmath ${X}$})|{\mathcal{D}_n}))^2,  \notag
\end{align}
where $g(\mbox{\boldmath ${r}$}|{\mathcal{D}_n})$ is the conditional density
function of $\mbox{\boldmath ${r}$}|{\mathcal{D}_n}$. Due to the complicated
form of $g(\mbox{\boldmath ${r}$}|{\mathcal{D}_n})$, integrations involved
in the above equations
are intractable. An alternative way is to use MCMC, but it is
computationally too demanding. In this paper, an h-likelihood method is
proposed to overcome this problem.

To implement the h-likelihood method, it is necessary to estimate the
unknown covariance kernel $k_i(\cdot,\cdot)$. We choose a covariance kernel
from a function family such as a squared exponential kernel or Mat\'{e}rn
class kernel. For each group, we can use different covariance kernels, for
example,
\begin{align}
k_i(\mbox{\boldmath ${u}$},\mbox{\boldmath ${v}$})=k(\mbox{\boldmath ${u}$},%
\mbox{\boldmath ${v}$};\mbox{\boldmath ${\theta}_i$}) =\theta _{i0}\exp {%
\left( -\frac{1}{2}\sum_{l=1}^{p}\eta _{il}(u_{l}-v_{l})^{2}\right) }%
+\sum_{l=1}^{p}\xi_{il}u_{l}v_{l},  \label{kerfun}
\end{align}
where $\mbox{\boldmath ${\theta}$}_i=\{\theta _{i0},\eta_{il},\xi
_{il},l=1,...,p\}$ are a set of parameters.

Let $\mbox{\boldmath ${\beta}$}=(\mbox{{\boldmath ${\theta}$}$_{1}^{T}$},...,%
\mbox{{\boldmath ${\theta}$}$_{I}^{T}$},\phi _{1},...,\phi _{I},\nu _{0},\nu
_{1})^{T}$. We propose the h-likelihood for the process regression model as
follows,
\begin{equation}
h_{0}=\sum_{i=1}^I\{\log (f_{\phi_i}(\mbox{{\boldmath ${y}$}$_{i}$}|%
\mbox{{\boldmath ${r}$}$_{i}$},f_i,\mbox{{\boldmath ${X}$}$_{i}$}))+\log
(f_{\theta_i}(\mbox{{\boldmath ${f}$}$_{in}$}|r_{i0},%
\mbox{{\boldmath
${X}$}$_{i}$}))\}+\log (f_{\nu _{0},\nu _{1}}(\mbox{\boldmath ${r}$})),
\notag
\end{equation}
where $f_{\phi_i}(\mbox{{\boldmath ${y}$}$_{i}$}|%
\mbox{{\boldmath
${r}$}$_{i}$} ,f_i,\mbox{{\boldmath ${X}$}$_{i}$})$ and $f_{\theta_i }(%
\mbox{{\boldmath ${f}$}$_{in}$}|r_{i0},\mbox{{\boldmath ${X}$}$_{i}$})$ are
the density functions of \newline
$N\left(\mbox{{\boldmath ${b}$}$_{J}$}\bigotimes
\mbox{{\boldmath
${f}$}$_{in}$},~\phi_i\mbox{{\boldmath ${r}$}$_{ei}$}\bigotimes
\mbox{{\boldmath
${I}$}$_{n}$}\right)$ and $N(0,~r_{i0}\mbox{{\boldmath ${K}$}$_{in}$})$,
respectively, and $f_{\nu _{0},\nu _{1}}(\mbox{\boldmath
${r}$})$ is the density function of $\mbox{\boldmath ${r}$}$.

By solving $\partial h_0/\partial \mbox{{\boldmath ${f}$}$_{in}$}=0$, we
obtain an estimate of $\mbox{{\boldmath ${f}$}$_{in}$}$,
\begin{align}  \label{condi.mean}
\tilde{\mbox{\boldmath ${f}$}}_{in}=\hat{\mbox{\boldmath ${f}$}}_{in}(%
\mbox{\boldmath ${\beta}$},\mbox{\boldmath ${r}$})=\left(\sum_{j=1}^J\frac{1%
}{r_{ij}}\mbox{{\boldmath ${I}$}$_{n}$}+\frac{\phi_i}{r_{i0}}
\mbox{{\boldmath
${K}$}$_{in}^{-1}$}\right)^{-1}\sum_{j=1}^J\frac{1}{r_{ij}}
\mbox{{\boldmath
${y}$}$_{ij}$}.
\end{align}
We can show that $\tilde{\mbox{\boldmath ${f}$}}_{in}=E(%
\mbox{{\boldmath
${f}$}$_{in}$}|\mbox{\boldmath ${r}$},\mathcal{D}_n)$. Thus, $\tilde{%
\mbox{\boldmath ${f}$}}_{in}$ is a BLUP (best linear unbiased prediction) of
$f_i(\mbox{{\boldmath ${X}$}$_{i}$})$ given $\mbox{\boldmath ${\beta}$}$ and
$\mbox{\boldmath
${r}$}$. Since $\mbox{\boldmath ${r}$}$ and $\mbox{\boldmath ${\beta}$}$ are
unknown, we need to estimate them. For $\mbox{\boldmath ${r}$}$, integrating
$h_{0}$ over $f_i$, $i=1,...,I$, we have
\begin{equation*}
h_{1}=\log \int \exp {(h_{0})}df_1\cdots df_I=\sum_{i=1}^{I}\{\log
(f_{\phi_i,\theta_i}(\mbox{{\boldmath ${y}$}$_{i}$}|%
\mbox{{\boldmath
${r}$}$_{i}$},\mbox{{\boldmath ${X}$}$_{i}$}))\}+\log (f_{\nu _{0},\nu _{1}}(%
\mbox{\boldmath ${r}$})),
\end{equation*}
where $f_{\phi_i,\theta_i}(\mbox{{\boldmath ${y}$}$_{i}$}|%
\mbox{{\boldmath
${r}$}$_{i}$},\mbox{{\boldmath ${X}$}$_{i}$})$ is the density function of $%
N(0,~\mbox{{\boldmath ${C}$}$_{ri}$})$. Maximizing $h_{1}$ over $%
\mbox{\boldmath ${r}$}$, we have the score equations,
\begin{align}
& \frac{\partial h_{1}}{\partial r_{i0}}=\frac{1}{2}{\mbox{Trace}}\left\{ %
\Big(\mbox{{\boldmath ${\alpha}$}$_{i}$}%
\mbox{{\boldmath
${\alpha}$}$_{i}^{T}$}-\mbox{{\boldmath ${C}$}$_{ri}^{-1}$}\Big)\frac{%
\partial \mbox{{\boldmath ${C}$}$_{ri}$}}{\partial r_{i0}}\right\} -\frac{%
\nu _{0}+1}{r_{i0}}+\frac{\nu _{0}-1}{r_{i0}^{2}}=0,  \label{r0score} \\
& \frac{\partial h_{1}}{\partial r_{ij}}=\frac{1}{2}{\mbox{Trace}}\left\{ %
\Big(\mbox{{\boldmath ${\alpha}$}$_{i}$}%
\mbox{{\boldmath
${\alpha}$}$_{i}^{T}$}-\mbox{{\boldmath ${C}$}$_{ri}^{-1}$}\Big)\frac{%
\partial \mbox{{\boldmath ${C}$}$_{ri}$}}{\partial r_{ij}}\right\} -\frac{%
\nu _{1}+1}{r_{ij}}+\frac{\nu _{1}-1}{r_{ij}^{2}}=0,j=1,...,J,
\label{rescore}
\end{align}
where $\mbox{\boldmath ${\alpha}_i$}=\mbox{{\boldmath${{C}}$}$_{ri}^{-1}$}%
\mbox{{\boldmath ${y}$}$_{i}$}$. The above score equations give an estimate
of $\mbox{{\boldmath ${r}$}$_{i}$}$, denoted by $\hat{\mbox{\boldmath ${r}$}}%
_{i}=(\hat{r}_{i0},...,\hat{r}_{iJ})^{T}$, $i=1,...,I$.

For $\mbox{\boldmath ${\beta}$}$, we use an adjusted profile likelihood,
\begin{equation*}
m=p_{r}(h_{1})=\sum_{i=1}^{I}\{\log (f_{\phi _{i},\theta _{i}}(%
\mbox{{\boldmath ${y}$}$_{i}$}|\hat{\mbox{\boldmath ${r}$}}_{i},%
\mbox{{\boldmath ${X}$}$_{i}$}))\}+\log (f_{\nu _{0},\nu _{1}}(\hat{%
\mbox{\boldmath ${r}$}}))-\frac{1}{2}\log |\mbox{\boldmath ${B}$}/(2\pi )|,
\end{equation*}
where
\begin{equation}
\mbox{\boldmath ${B}$}=-\frac{\partial ^{2}h_{1}}{\partial
\mbox{\boldmath
${r}$}\partial \mbox{{\boldmath ${r}$}$^{T}$}}\Big|_{\mbox{\boldmath ${r}$}=%
\hat{\mbox{\boldmath ${r}$}}}.  \notag
\end{equation}
The adjusted profile likelihood $m$ is the Laplace approximation to the
integrated likelihood $\log \int \exp (h_{1})d\mbox{\boldmath ${r}$}$ $%
=\sum_{i=1}^I\log (f_{\phi_i,\theta_i}(\mbox{{\boldmath ${y}$}$_{i}$}|%
\mbox{{\boldmath ${X}$}$_{i}$}))$. This leads to a score equation for $%
\mbox{\boldmath ${\beta}$}$
\begin{align}
\frac{\partial m}{\partial \beta _{l}}=& \frac{1}{2}{\mbox{Trace}}\left\{ %
\Big(\mbox{{\boldmath ${\alpha}$}$_{i}$}%
\mbox{{\boldmath
${\alpha}$}$_{i}^{T}$}-\mbox{{\boldmath ${C}$}$_{ri}^{-1}$}\Big)\frac{%
\partial \mbox{{\boldmath ${C}$}$_{ri}$}}{\partial \beta _{l}}\right\} -%
\frac{1}{2}{\mbox{Trace}}\left( \mbox{{\boldmath ${B}$}$^{-1}$}\frac{%
\partial \mbox{\boldmath ${B}$}}{\partial \beta _{l}}\right)  \notag \\
& +\sum_{i=1}^I\sum_{j=0}^{J}\frac{\partial m}{\partial r_{j}}\frac{\partial
r_{j}}{\partial \beta _{l}}=0,  \label{betascore}
\end{align}
where $\beta _{l}$ is the $l$th element of $\mbox{\boldmath ${\beta}$}$.
Maximizing $m$ with respect to $\mbox{\boldmath ${\beta}$}$, we have an
estimate of $\mbox{\boldmath ${\beta}$}$, denoted by $\hat{%
\mbox{\boldmath
${\beta}$}}$.

From (\ref{condi.mean}) we have $\hat{\mbox{\boldmath ${f}$}}_{in}=\hat{%
\mbox{\boldmath ${f}$}}_{in}(\hat{\mbox{\boldmath ${\beta}$}},\hat{%
\mbox{\boldmath ${r}$}})$. From (\ref{r0score})-(\ref{betascore}), we see
that the score equations for $\mbox{\boldmath ${r}$}$ and $%
\mbox{\boldmath
${\beta}$}$ are even invariant in $\mbox{\boldmath
${y}$}$, which leads to even invariant forms for $\hat{\mbox{\boldmath ${r}$}%
}$ and $\hat{\mbox{\boldmath ${\beta}$}}$. Using results in \cite{r2}, we
can show that the estimate $\hat{\mbox{\boldmath ${f}$}}_{in}$ is unbiased.
Plugging estimates of $\mbox{\boldmath ${r}$}$ and $%
\mbox{\boldmath
${\beta}$}$ in (\ref{cond.cov}), it gives an estimate of the variance of $%
\hat{\mbox{\boldmath ${f}$}}_{in}$. But generally (\ref{cond.cov})
underestimates the variance of $\hat{\mbox{\boldmath ${f}$}}_{in}$ because
it does not take into account the variance increase caused by estimating
unknown parameters. We use the following procedure to improve the estimate.
Denote the inverse of the negative Hessian matrix of $h_{0}$ with respect to
$\mbox{{\boldmath ${f}$}$_{in}$}$ and $\mbox{{\boldmath
${r}$}$_{i}$}$ by $\mbox{{\boldmath ${H}$}$_{in}$}$. The first prime $%
n\times n$ submatrix of $\mbox{{\boldmath ${H}$}$_{in}$}$ is used as an
estimate of the variance of $\hat{\mbox{\boldmath ${f}$}}_{in}$.

For a new point $\mbox{\boldmath ${z}$}$ at the $i$th group, replacing $%
\mbox{\boldmath ${r}$} $ and $\mbox{\boldmath ${\beta}$}$ with $\hat{%
\mbox{\boldmath ${r}$}}$ and $\hat{\mbox{\boldmath ${\beta}$}}$ in (\ref
{cond.pred}) gives an estimate of $f_{i}(\mbox{\boldmath ${z}$})$, saying $%
\hat{f}_{i}(\mbox{\boldmath ${z}$})$. Similar to $h_{0}$, we can derive an
h-log likelihood function with respect to $y_{i}(\mbox{\boldmath ${z}$})$, $%
f_{i}(\mbox{\boldmath ${z}$})$, $\mbox{{\boldmath ${y}$}$_{i}$}$ and $%
\mbox{{\boldmath ${r}$}$_{i}$}$, denoted by $h_{z}$. Here $y_{i}(%
\mbox{\boldmath ${z}$})$ is also unobservable. Then, the inverse of the
negative Hessian matrix of $h_{z}$ with respect to $y_{i}(%
\mbox{\boldmath
${z}$})$, $f_{i}(\mbox{\boldmath ${z}$})$, $\mbox{{\boldmath
${f}$}$_{in}$}$ and $\mbox{{\boldmath ${r}$}$_{i}$}$ is computed, saying $%
\mbox{{\boldmath ${H}$}$_{iz}$}$, and $y_{i}(\mbox{\boldmath ${z}$})$ is
replaced with $\hat{f}_{i}(\mbox{\boldmath ${z}$})$ and other unknown items
are replaced with their estimates. The first diagonal component of $%
\mbox{{\boldmath ${H}$}$_{iz}$}$ is taken as an estimate of the variance of $%
\hat{f}_{i}(\mbox{\boldmath ${z}$})$.

\textbf{Remark 3.} From (\ref{condi.mean}), random effect $r_{ij}$ can be
used to detect outlying curves. For example, if the $j$th curve in group $i$
is outlying (having large errors), then $r_{ij}$ may have a large value to
give smaller weight to the response which can reduce influence of an
outlying curve on predictor of $f_{i}$. Thus, $\hat{r}_{ij}$ may be used as
an indicator to find outlying curves. The detailed discussion will be given
in Section 4. 

\subsection{Information consistency}

Suppose that for each group $i$, there are $J$ curves. Let $p_{\phi _{0i}}(%
\mbox{{\boldmath ${y}$}$_{i}$}|f_{0i},\mbox{{\boldmath ${X}$}$_{i}$})$ be
the density function to generate the data $\mbox{{\boldmath ${y}$}$_{i}$}$
given $\mbox{{\boldmath ${X}$}$_{i}$}$ under the true model (\ref{true}),
where $f_{0i}$ is the true underlying function of $f_i$ and $\phi_{0i}$ is
the true value of $\phi_i$. Let $p_{\theta_i,r_i }(f)$ be a measure of
random process $f$ on space ${\mathcal{F}}=\{f(\cdot):\mathcal{X}\rightarrow
R\}$ for given random effect $\mbox{{\boldmath ${r}$}$_{i}$}$. Let
\begin{equation*}
p_{\phi_i ,\theta_i,r_i}(\mbox{{\boldmath ${y}$}$_{i}$}|%
\mbox{{\boldmath
${X}$}$_{i}$})=\int_{{\mathcal{F}}}p_{\phi_i,r_i }(%
\mbox{{\boldmath
${y}$}$_{i}$}|f,\mbox{{\boldmath ${X}$}$_{i}$})dp_{\theta_i,r_i }(f),
\end{equation*}
be the density function to generate the data $\mbox{{\boldmath ${y}$}$_{i}$}$
given $\mbox{{\boldmath ${X}$}$_{i}$}$ under the assumed model (\ref{assumed}%
) and given $\mbox{{\boldmath ${r}$}$_{i}$}$. Thus, the assumed model (\ref
{assumed}) is not the same as the true underlying model (\ref{true}). Let $%
p_{\phi _{0i},\hat \theta_i, \hat r_i}(\mbox{{\boldmath ${y}$}$_{i}$}|%
\mbox{{\boldmath ${X}$}$_{i}$})$ be the estimated density function under the
assumed model (\ref{assumed}), where $\hat{\mbox{\boldmath ${\theta}$}}_{i}$
and $\hat{\mbox{\boldmath ${r}$}}_{i}$ are the estimators of parameter $%
\mbox{{\boldmath ${\theta}$}$_{i}$}$ and $\mbox{{\boldmath ${r}$}$_{i}$}$.
Denote $D[p_{1},p_{2}]=\int (\log {p_{1}}-\log {p_{2}})dp_{1}$ by the
Kullback-Leibler distance between two densities $p_{1} $ and $p_{2}$.

Following \cite{r18}, $\hat{\mbox{\boldmath ${\theta}$}}_{i}$ and $\hat{%
\mbox{\boldmath ${r}$}}_{i}$ are consistent estimators of $%
\mbox{{\boldmath
${\theta}$}$_{i}$}$ and $\mbox{{\boldmath ${r}$}$_{i}$}$, respectively. Then
we have the next theorem (the proof is given in Appendix).

\vskip10pt \noindent \textbf{Theorem 1} \textit{Under the appropriate
conditions in Appendix, for each group $i$, we have
\begin{equation*}
\frac{1}{n} E_{\mbox{{\boldmath ${X}$}$_{i}$}}(D[p_{\phi _{0i}}(%
\mbox{{\boldmath ${y}$}$_{i}$}|f_{0i},\mbox{{\boldmath ${X}$}$_{i}$}%
),p_{\phi _{0i},\hat{\theta}_i,\hat r_i}(\mbox{{\boldmath ${y}$}$_{i}$}|%
\mbox{{\boldmath ${X}$}$_{i}$})])\longrightarrow 0,{\mbox{as}}~~n\rightarrow
\infty ,
\end{equation*}
where the expectation is taken over the distribution of $%
\mbox{{\boldmath
${X}$}$_{i}$}$. }

\vskip 10pt

Theorem 1 shows that the Kullback-Leibler distance between two density
functions for $\mbox{{\boldmath ${y}$}$_{i}$}|\mbox{{\boldmath ${X}$}$_{i}$}$
from the true and the assumed models tends to zero asymptotically. Similar
to \cite{r11}, Theorem 1 is called information consistency which was first
proved for GPR in \cite{r15} where they used Bayesian prediction strategies
to derive prediction distribution of $f_i$ conditional on the observed data.

\section{Numerical studies}

\subsection{Simulation studies}

Numerical studies were conducted to evaluate performance of the four models:
GP-GP (GPR), GP-TP, TP-TP and TP-GP models in Section 2. From the numerical
studies, we find that TP-TP and GP-TP models behave similarly, so do TP-GP
and GP-GP models. So only results of GPR (GP-GP) and GP-TP models are
presented in this subsection. We take $n=10$. Data are generated from the
process model (\ref{assumed}) where $f_{i}$ follows a GP with mean $0$ and
the covariance kernel (\ref{kerfun}), and error term follows two different
distributions: Gaussian distribution with mean 0 and variance $\phi _{i}$,
and extended T process $ETP(\nu ,k_{\epsilon })$ with $k_{\epsilon
}(u,v)=\phi I(u=v)$. The $30$ points evenly spaced in [0,~3.0] are generated
for covariate, denoted by $S$. We take $n$ points with orders evenly spaced
in $S$ as training data, and the left as test data. To study robustness of
the proposed methods, responses from the $6$th curve in each group of the
training data are added with extra errors: constant error $\gamma $ or
random error $t_{2}+\gamma $ where $t_{2}$ follows the Student
t-distribution with degrees of freedom 2. We take $\gamma =0.5$, 1.0 and
2.0. Prediction performance is measured by mean squared error $%
MSE=\sum_{i=1}^{I}\sum_{k=1}^{m}(\hat{f}_{i}(x_{k}^{\ast
})-f_{0i}(x_{k}^{\ast }))^{2}/(nI)$, where $\{x_{k}^{\ast }:k=1,...,m\}$ are
the test data points. All results are based on 500 replications.

To compare the proposed models with the joint erorr model (eTPR) in \cite
{r11}, we take $I=1$ and $J=6$, and add constant or random disturbances for
Gaussian error, and add constant one for ETP error. The true values of the
parameters are $\mbox{{\boldmath${\theta}$}$_{1}$}=(\theta _{10},\eta
_{11},\xi _{11})=(0.1,10,0.1)$, $\phi_{1}=\phi=0.2$ and $\nu=1$ or 2. MSEs
of the predictions using the GPR, eTPR (joint error model) and GP-TP
(independent error model) are presented in Table~\ref{tab1-1}. It shows that
predictions from the proposed method GP-TP has the smallest MSE, while eTPR
behaves similar to GPR because of the large sample sizes (see the discussion
in Remark 1). For the cases with large constant disturbance, random
disturbance, or small value of $\nu$, GP-TP performs particularly better
than the other two models.

\begin{center}
******Insert Table \ref{tab1-1} here*****
\end{center}

\begin{figure}[h!]
\begin{center}
\includegraphics[height=0.8\textwidth,width=0.9\textwidth]{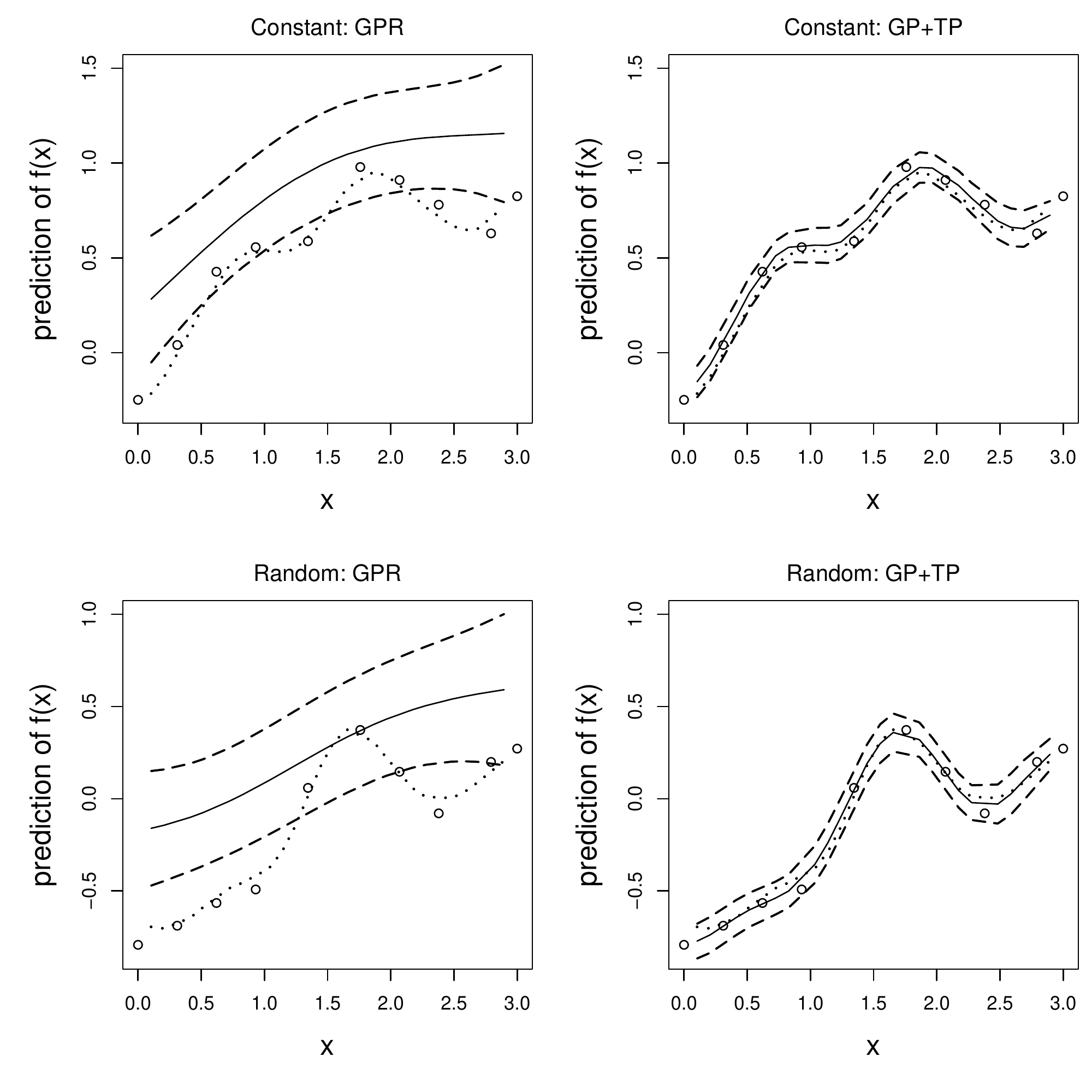}
\end{center}
\caption{Prediction curves from GPR (the 1st column) and GP-TP (the 2nd
column) for data with constant and random disturbances to the 6-th curve,
where the circles represent the means of the observed data excluding the
6-th curve, the dotted line stands for the true curve, and solid and dashed
lines stand for predicted curves and their 95\% confidence bounds. }
\label{fig2}
\end{figure}

Now we study performance of the GPR (GP-GP) and GP-TP for batch data with
more than one group. We take $I=2$ and $J=6$. Figure \ref{fig2} plots
predictive curves for two sets of simulated data with the constant
disturbance of $2.0$ and the random one of $t_{2}+2.0$ respectively. The
values of the other parameters are $\phi _{1}=\phi _{2}=0.01$, $%
\mbox{{\boldmath
${\theta}$}$_{i}$}=(\theta _{i0},\eta _{i1},\xi _{i1})=(0.1,5,0.1),i=1,2$.
The upper panel presents the results from the models with constant random
disturbances, and the lower panel for the random ones. The disturbance is
added to the 6th curve and thus it is an outlying curve. The means for
observed data points excluding the 6th curve are computed and represented by
circles in the figure.
The dotted line stand for the true curve, solid and dashed lines stand for
the predicted curve and their 95\% point-wise confidence bounds. We see that
the prediction from the GP-TP is much closer to the true curve than that
from the GPR, indicating
that the GP-TP is more robust against outlying curves compared to the GPR.

Table \ref{tab1} lists simulation study results of MSEs from the two models
based on 500 replications, where $\phi _{1}=\phi _{2}=0.2$, $%
\mbox{{\boldmath
${\theta}$}$_{i}$}=(\theta _{i0},\eta _{i1},\xi _{i1})=(0.1,10,0.1),i=1,2$.
It shows that the GP-TP method has smaller value of MSE than the GPR.
Especially, the GP-TP performs much better than the GPR for large values of $%
\gamma$ (1.0 and 2.0). 

\begin{center}
******Insert Table \ref{tab1} and Figure \ref{fig2} here*****
\end{center}

Estimates of the random effects from the GP-TP are presented in Table~\ref
{tab2}. For the example, there are six independently observed curves and
thus six random effects in each group, denoted by $r_{i1}$ to $r_{i6}$, $%
i=1,2$. We see that $r_{i6}$, which is corresponding to the outlying curve,
has much larger value than the others, especially for $\gamma =1.0$ and $2.0$%
, i.e. large values of $\gamma$. The remaining, $r_{ij}$, $j=1,...,5$, have
similar values. From (\ref{condi.mean}), we know that bigger random effects
give smaller weight to the corresponding response curve. Hence, the proposed
GP-TP is more robust against outlying curves compared to the GPR. In
practice, the values of estimated random effects can be used to detect
outlying curves, such as the $\hat{r}_{i6}$ shown in Table \ref{tab2}.

\begin{center}
******Insert Table \ref{tab2} here*****
\end{center}

\subsection{Real example}

Motor learning can be assessed more quickly and robustly than outcomes from
rehabilitation. \cite{Davison14} proposed to utilize a commodity input
device to play a bespoke video game to measure the critical components of
motor learning. To detect how simple changes in therapist instruction change
motor performance and learning, experiment has been conducted by either
giving a single objective (single instruction) or by breaking the task down
into its two sequential action phases (double instruction). High
spatial-temporal resolution data are recorded when participants play a
bespoke video game. We then calculate the mean distance between the target
and the avatar during the lock and track phase. This index reflects
predominantly feedback mechanisms and error correction.

The game data consists of two datasets: 24 young persons and 26 old adults.
For each dataset, one half of the subjects received single instruction,
saying single treatment group, and the others had twice, denoted by double
treatment group. There are $n=180 $ mean distances (meandist) recorded for
all subjects. The effect of instruction is studied for young and old adult
persons, respectively. For each dataset, we separately have two groups,
denoted by young with single instruction (young-single), young with double
instruction (young-double), old adults with single instruction (old-single),
and old adults with double instruction (old-double).

The estimates of random effects in the GP-TP model, $\{r_{i},i=1,...J\}$,
are presented in Table~\ref{tab4}. We find the estimated random effect for
the 4th subject in the young-double group has a much larger value compared
to the others, so do the 13th subject in the old-single group. Consequently
this model gives small weight to those possible outlying curves and thus
reduce their influence on prediction. Table \ref{tab4} also lists the
estimation of random effects after those two curves are deleted ( namely
`double-4' and `single-13'), showing that the estimation seem to be more
regular now.

\begin{center}
******Insert Table \ref{tab4} here*****
\end{center}

\begin{figure}[h!]
\begin{center}
\includegraphics[height=0.9\textwidth,width=0.9\textwidth]{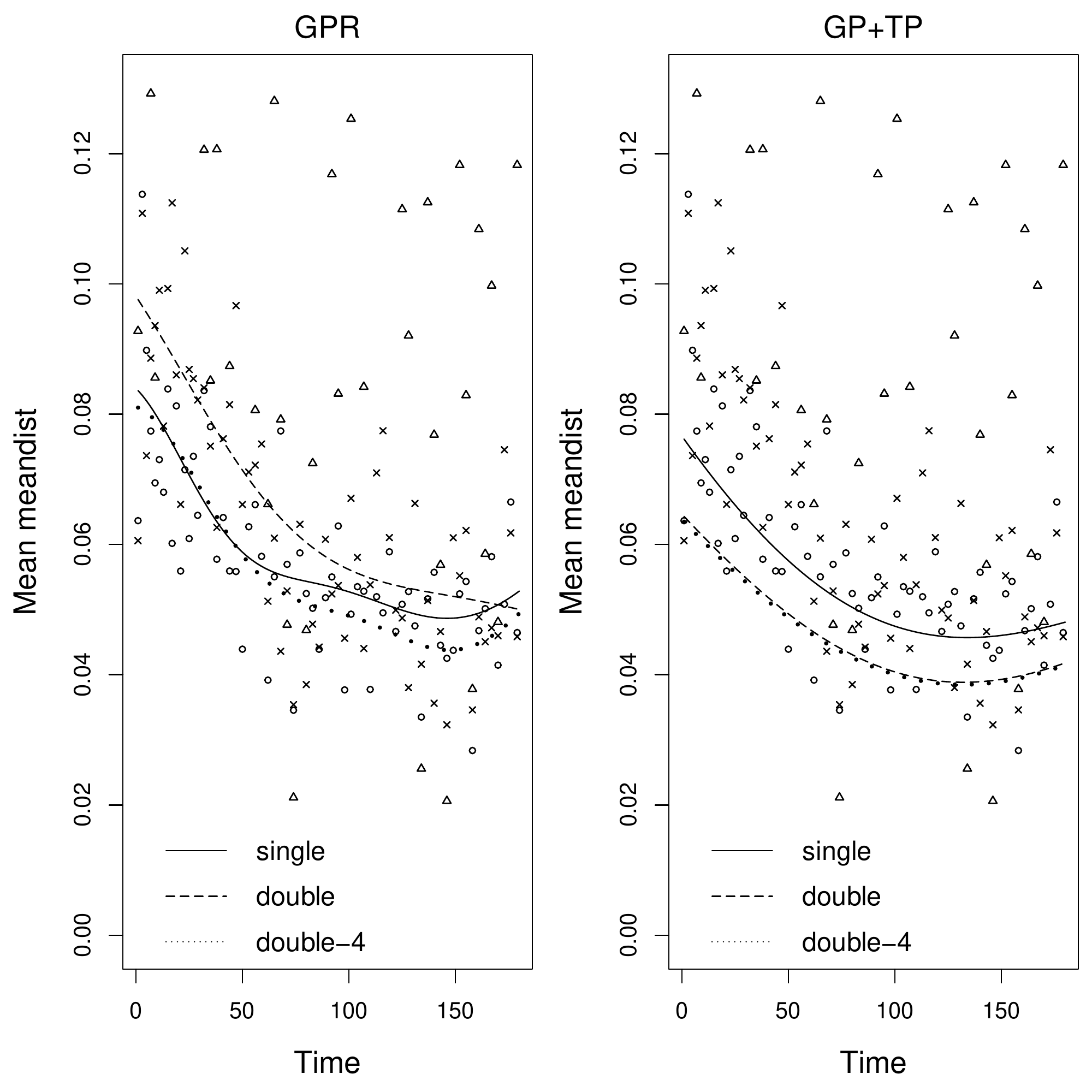}
\end{center}
\caption{Prediction curves for the young person dataset from GPR and GP-TP
models. Cross and circle points represent average values of response
meandist of single and double instruction groups respectively, and triangle
point stands for meandist of the 4th subject in double instruction group.
The solid line stands for the predictive curve of meandist for single
instruction group. The dashed and dotted lines stand for the predictions of
meandist for double instruction group with and without the 4th subject
respectively. }
\label{fig4}
\end{figure}

\begin{figure}[h!]
\begin{center}
\includegraphics[height=0.9\textwidth,width=0.9\textwidth]{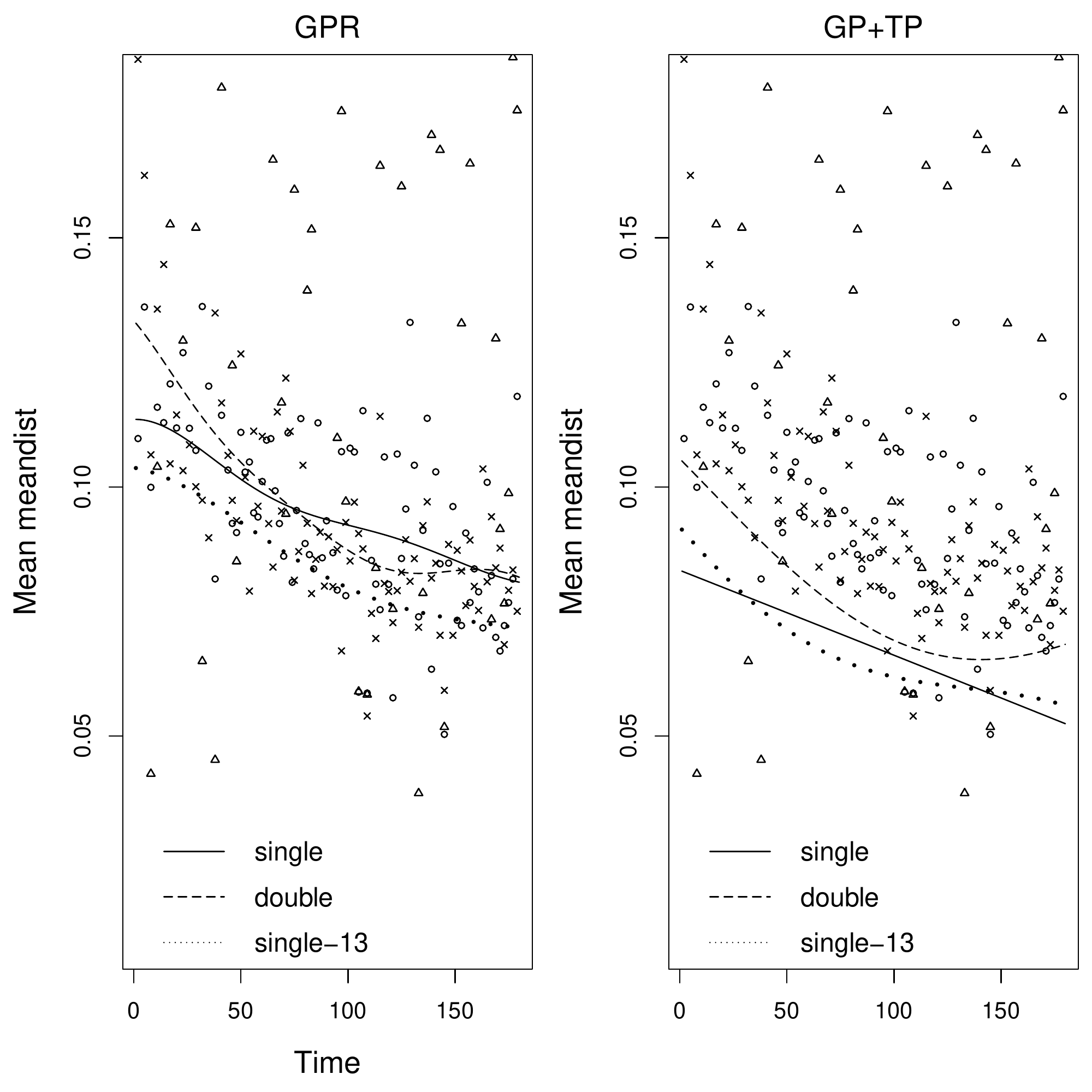}
\end{center}
\caption{Prediction curves for the old adult dataset from GPR and GP-TP
models. Cross and circle points represent average values of response
meandist of single and double instruction groups respectively, and triangle
point stands for meandist of the 4th subject in double instruction group.
The solid and dotted lines stand for the predictive curve of meandist for
single instruction group with and without the 13th subject respectively. The
dashed line stands for the predictions of meandist for double instruction
group. }
\label{fig5}
\end{figure}

Prediction curves from the two models are respectively plotted in Figures
\ref{fig4} and \ref{fig5} for young person and old adult datasets. In both
figures, cross and circle points represent average values of meandist for
single and double instruction groups respectively, and triangle point stands
for meandist of the 4th and 13th subject (outliers). We can see that the 4th
subject has larger meandist than the others in young-double subgroup,
implying that it may be an outlying curve. The prediction curves in Figure
\ref{fig4} shows that the GP-TP model is almost not affected by the 4th
subject, while the affection to the GPR model is quite significant. When the
4th subject is not included in the data set, the predicted values for the
double group calculated from the GPR are almost the same as the ones from
the single group in the area of the first half, but the GP-TP model shows
the difference uniformly in the whole area no matter whether the 4th subject
is included or not. This shows the property of robustness of the GP-TP
model. Figure \ref{fig5} shows a similar result.

\begin{center}
******Insert Figures \ref{fig4} and \ref{fig5} here*****
\end{center}

\section{\textbf{Concluding remarks}}

This paper develops a robust estimation procedure via h-likelihood for an
independent error model with functional batch data. The estimated random
effects are useful to detect outlying curves. Unbiasness and information
consistency are shown. Numerical studies show that the proposed estimation
procedure is robust against outlying curves, and has a better performance in
the presence of outliers compared to the GPR and eTPR models. We focused our
discussion in this paper on the TP+TP model, but the estimation procedure
can be applied straightly to other types of models, for example, functional
regression models with independent errors of multivariate t distribution
(MVT). In this case, model (\ref{assumed}) is modified as: $f_{i}\sim
GP(k_{i})$, $\epsilon _{ijk}\sim MVT(\nu _{1},\phi _{i})$, for $%
i=1,...,I,j=1,...,J,k=1,...,n$, where $f_{i}$ and $\epsilon_{ijk}$ are
independent. The errors can be defined equivalently to
\begin{align}
&\epsilon_{ijk}|r_{ijk}\sim N(0,r_{ijk}\phi_i),~~r_{ijk}\sim \mathrm{{IG}%
(\nu_1 ,\nu_1 -1)}.  \notag
\end{align}
Thus we can use the h-likelihood method given in Section 3 to estimate the
unknown function $f_i$.

In addition, the proposed method can be extended to generalized linear model
with functional data.

\section*{Acknowledgement}

Wang's work is supported by funds of the State Key Program of National
Natural Science of China (No. 11231010) and National Natural Science of
China (No. 11471302). Lee's work is funded by an NRF grant of Korea
government (MEST) (No. 2011-0030810) and Science Original Technology
Research Program for Brain Science of Ministry of Science, ICT and Future
Planning (NRF-2014M3C7A1062896).

\section*{\textbf{Appendix}}

\setcounter{equation}{0}
\renewcommand\theequation{A.\arabic {equation} }

Hereafter, let $c$ be any positive constant independent of $n$, which may

stands for various

values in different places. Without loss of generality, firstly we set $%
I=J=1 $ which is the situation in Section 2. Let $p_{\phi _{0}}(%
\mbox{{\boldmath
${y}$}}|f_{0},\mbox{{\boldmath
${X}$}})$ be the density function to generate the data $%
\mbox{{\boldmath
${y}$}}$ given $\mbox{{\boldmath
${X}$}}$ under the true model (\ref{model2-1}), where $f_{0}$ is the true
underlying function of $f$. Let $p_{\theta,r }(f)$ be a measure of random
process $f$ on space ${\mathcal{F}}=\{f(\cdot):\mathcal{X}\rightarrow R\}$
for given random effect $\mbox{\boldmath ${r}$}$. Let
\begin{equation*}
p_{\phi ,\theta,r}(\mbox{{\boldmath ${y}$}}|\mbox{{\boldmath
${X}$}})=\int_{{\mathcal{F}}}p_{\phi,r }(\mbox{{\boldmath
${y}$}}|f,\mbox{{\boldmath
${X}$}})dp_{\theta,r }(f),
\end{equation*}
be the density function to generate the data $\mbox{{\boldmath ${y}$}}$
given $\mbox{{\boldmath
${X}$}}$ under the assumed model (\ref{model2-2}) and given $%
\mbox{\boldmath
${r}$}$. Here $\phi $ is the common parameter in both models and let $\phi_0$
be the true value of $\phi$. Let $p_{\phi _{0},\hat \theta, \hat r}(%
\mbox{{\boldmath
${y}$}}|\mbox{{\boldmath ${X}$}})$ be the estimated density function under
the assumed model (\ref{model2-2}). Before proving Theorem 1, we need the
following Lemma.

\vskip 10pt \noindent \textbf{Lemma 1} \textit{Suppose $%
\mbox{{\boldmath
${y}$}}=\{y_{1},...,y_{n}\}$ are generated from model (\ref{model2-2}) with
the mean function $h(\mbox{\boldmath ${x}$})=0$, and covariance kernel
function $k$ is bounded and continuous in parameter $%
\mbox{\boldmath
${\theta}$}$. It also assumes that the estimate $\hat{%
\mbox{\boldmath
${\beta}$}}$ and $\hat{\mbox{\boldmath ${r}$}}$ are consistent estimators of
$\mbox{\boldmath ${\beta}$}$ and $\mbox{\boldmath ${r}$}$, respectively.
Then for any $\varepsilon>0$, when $n$ is large enough, we have
\begin{align}
&\frac{1}{n}(-\log p_{\phi_0,\hat\theta,\hat r}(\mbox{{\boldmath ${y}$}$$}|%
\mbox{{\boldmath ${X}$}$$})+\log p_{\phi_0}( \mbox{{\boldmath
${y}$}$$}|f_{0},\mbox{{\boldmath ${X}$}$$}))  \notag \\
\leq& \frac{1}{2n}\Big\{\int \log|\mbox{{\boldmath ${I}$}$_{n}$}+r_0 %
\mbox{{\boldmath ${K}$}$_{n}$}/(r_1\phi_0)|g_{\nu_0} (r_0) g^*_{\nu_1} (r_1)
dr_0 dr_1+||f_{0}||^2_k+c\Big\}+ \varepsilon,  \notag
\end{align}
where $\mbox{{\boldmath ${K}$}$_{n}$}=(k(\mbox{{\boldmath ${x}$}$_{j}$},%
\mbox{{\boldmath ${x}$}$_{l}$}))_{n\times n}$, $%
\mbox{{\boldmath
${I}$}$_{n}$}$ is the $n\times n$ identity matrix, $||f_{0}||_k$ is the
reproducing kernel Hilbert space norm of $f_{0}$ associated with kernel
function $k(\cdot,\cdot;\mbox{\boldmath ${\theta}$})$, ${g}_{\nu_0}(\cdot)$
and ${g}_{\nu_1}^*(\cdot)$ are the density functions of $IG(\nu_0,\nu_0-1)$
and $IG(\nu_1+n/2,(\nu_1-1)+q^2/2)$, respectively, and $q^2=( %
\mbox{{\boldmath ${y}$}$$}-f_{0}(\mbox{{\boldmath ${X}$}$$}))^T( %
\mbox{{\boldmath ${y}$}$$}-f_{0}(\mbox{{\boldmath ${X}$}$$}))/\phi_0$. }

\vskip 10pt \textbf{Proof}: Suppose that for any given $%
\mbox{\boldmath
${r}$}=(r_0,r_1)^T$, we have
\begin{align}  \label{Aeq1}
&-\log p_{\phi_0,\theta, r}(\mbox{{\boldmath ${y}$}$$}|%
\mbox{{\boldmath
${X}$}$$})+\log p_{\phi_0, r}( \mbox{{\boldmath
${y}$}$$}|f_{0},\mbox{{\boldmath ${X}$}$$}))  \notag \\
\leq& \frac{1}{2}\log|\mbox{{\boldmath ${I}$}$_{n}$}+r_0
\mbox{{\boldmath
${K}$}$_{n}$}/(r_1\phi_0)|+ \frac{r_0}{2}(||f_{0}||^2_k+c)+c+n\varepsilon.
\end{align}
Let $\hat{\mbox{\boldmath ${r}$}}$ and $\tilde{\mbox{\boldmath ${r}$}}$ be
maximizers of functions $\log p_{\phi_0, \theta, r}(\mbox{\boldmath ${y }$}
| \mbox{\boldmath ${X}$})+\log f_\nu (\mbox{\boldmath ${r}$})$ and $\log
p_{\phi_0, \theta, r}(\mbox{\boldmath ${y }$} | \mbox{\boldmath ${X}$})$,
respectively, where $f_\nu (\mbox{\boldmath ${r}$})=g_{\nu_0}(r_0)g_{%
\nu_1}(r_1)$. We have
\begin{align}
1\leq \frac{\log p_{\phi_0, \theta, \tilde r}(\mbox{\boldmath ${y }$} | %
\mbox{\boldmath ${X}$})}{\log p_{\phi_0, \theta, \hat r}(%
\mbox{\boldmath ${y
}$} | \mbox{\boldmath ${X}$})} \leq 1+\frac{\log f_\nu(\hat{%
\mbox{\boldmath
${r}$}})-\log f_\nu(\tilde{\mbox{\boldmath ${r}$}})}{\log p_{\phi_0, \theta,
\hat r}(\mbox{\boldmath ${y }$} | \mbox{\boldmath ${X}$})},  \notag
\end{align}
which indicates that
\begin{align}
p_{\phi_0, \theta, \tilde r}(\mbox{\boldmath ${y }$} | \mbox{\boldmath ${X}$}%
)\leq p_{\phi_0, \theta, \hat r}(\mbox{\boldmath ${y }$} |
\mbox{\boldmath
${X}$})^{1+\varepsilon},  \notag
\end{align}
because $f_\nu (\mbox{\boldmath ${r}$})$ is independent of $n$ while $\log
p_{\phi_0, \theta, \hat r}(\mbox{\boldmath ${y }$} | \mbox{\boldmath ${X}$})$
tends to infinity as $n$ goes to $\infty$. Thence,
\begin{align}
\int p_{\phi_0, \theta, r}(\mbox{\boldmath ${y }$} | \mbox{\boldmath ${X}$})
f_\nu (\mbox{\boldmath ${r}$}) d \mbox{\boldmath ${r }$} \leq p_{\phi_0,
\theta, \hat r}(\mbox{\boldmath ${y }$} | \mbox{\boldmath ${X}$}%
)^{1+\varepsilon}.  \label{Aeq2}
\end{align}

By simple computation, we show that
\begin{align}  \label{Aeq3}
&\int p_{\phi_0, r}(\mbox{\boldmath ${y }$} |f_0, \mbox{\boldmath ${X}$})
\exp\{-( \frac{1}{2}\log|\mbox{{\boldmath ${I}$}$_{n}$}+r_0
\mbox{{\boldmath
${K}$}$_{n}$}/(r_1\phi_0)|+\frac{r_0}{2}(||f_{0}||^2_k+c))\} f_\nu (%
\mbox{\boldmath ${r}$}) d\mbox{\boldmath ${r  }$}  \notag \\
=&p_{\phi_0}(\mbox{\boldmath ${y }$} |f_0, \mbox{\boldmath ${X}$})
\int\exp\{-( \frac{1}{2}\log|\mbox{{\boldmath ${I}$}$_{n}$}+r_0 %
\mbox{{\boldmath ${K}$}$_{n}$}/(r_1\phi_0)|+\frac{r_0}{2}(||f_{0}||^2_k+c))\}
\notag \\
& g_{\nu_0} (r_0) g^*_{\nu_1} (r_1) dr_0 dr_1.
\end{align}
From (\ref{Aeq1}), (\ref{Aeq2}) and (\ref{Aeq3}), we have
\begin{align}
&\frac{1}{n}(-\log p_{\phi_0,\theta,\hat r}(\mbox{{\boldmath ${y}$}$$}|%
\mbox{{\boldmath ${X}$}$$})+\log p_{\phi_0}( \mbox{{\boldmath
${y}$}$$}|f_{0},\mbox{{\boldmath ${X}$}$$}))  \notag \\
\leq& -\frac{1}{n}\log\Big\{\int\exp\{-( \frac{1}{2}\log|%
\mbox{{\boldmath
${I}$}$_{n}$}+r_0 \mbox{{\boldmath ${K}$}$_{n}$}/(r_1\phi_0)|+\frac{r_0}{2}%
(||f_{0}||^2_k+c))\}  \notag \\
&g_\nu (r_0) g^*_\nu (r_1) dr_0 dr_1\Big\}+\varepsilon  \notag \\
\leq & \frac{1}{2n} \Big\{\int \log|\mbox{{\boldmath ${I}$}$_{n}$}+r_0 %
\mbox{{\boldmath ${K}$}$_{n}$}/(r_1\phi_0)|g_\nu (r_0) g^*_\nu (r_1) dr_0
dr_1+ ||f_{0}||^2_k+c\Big\}+\varepsilon,  \notag
\end{align}
which shows that Lemma 1 holds.

Now let us prove the inequality (\ref{Aeq1}). Following proofs of Theorem 1
in \cite{r15} and Lemma 1 in \cite{r10}, it is sufficient to prove (\ref
{Aeq1}) when the true underlying function has the expression
\begin{align}
&{f}_{0}(\cdot)=r_0\sum_{l=1}^n\alpha_l k(\mbox{\boldmath ${x}$},%
\mbox{{\boldmath ${x}$}$_{i}$};\mbox{{\boldmath ${\theta}$}$_{}$})\doteq r_0
K(\cdot)\mbox{\boldmath ${\alpha}$},  \notag
\end{align}
where $K(\cdot)=(k(\mbox{\boldmath ${x}$},\mbox{{\boldmath ${x}$}$_{1}$};%
\mbox{{\boldmath ${\theta}$}$_{}$}),...,k(\mbox{\boldmath ${x}$},%
\mbox{{\boldmath
${x}$}$_{n}$};\mbox{{\boldmath ${\theta}$}$_{}$}))$ and $%
\mbox{\boldmath
${\alpha}$}=(\alpha_1,...,\alpha_n)^T\in R^n$.

Let $P$ be a measure induced by $GP(0,r_0k(\cdot,\cdot;\hat{%
\mbox{\boldmath
${\theta}$}}))$. Let $Q$ be the density function of normal distribution $%
N(f_0(\mbox{\boldmath ${X}$}), r_0\mbox{{\boldmath ${K}$}$_{n}$}(r_0%
\mbox{{\boldmath ${K}$}$_{n}$}/(r_1\phi_0)+\mbox{{\boldmath ${I}$}$_{n}$}%
)^{-1})$. We can show that $Q$ is the posterior distribution of $\tilde f$
from a model with prior $GP(0, r_0k(\cdot,\cdot;%
\mbox{{\boldmath
${\theta}$}$_{}$} ))$ and Gaussian likelihood term $\prod_{l=1}^n N(\hat
y_{i}|\tilde f( \mbox{{\boldmath ${x}$}$_{i}$}),r_1\phi_0)$, where $\hat{%
\mbox{\boldmath ${y}$}}=(\hat y_{1},...,\hat y_{n})^T=(r_0%
\mbox{{\boldmath
${K}$}$_{n}$}+r_1\phi_0\mbox{{\boldmath
${I}$}$_{n}$})\mbox{\boldmath ${\alpha}$}$. Then we have $E_{Q}(\tilde
f)=f_{0}$, where the expectation is taken under probability density $Q$.
From Fenchel-Legendre duality relationship in \cite{r16} and \cite{r17}, we
have
\begin{align}  \label{FLD}
-\log p_{\phi_0,\theta, r}(\mbox{{\boldmath ${y}$}$$}|%
\mbox{{\boldmath
${X}$}$$}) \leq E_{Q}(-\log p_{\phi_0, r}(\mbox{{\boldmath ${y}$}$$}| \tilde{%
f}, \mbox{{\boldmath ${X}$}$$})+D[Q,P].
\end{align}
Let $\mbox{\boldmath ${B}$}=\mbox{{\boldmath ${I}$}$_{n}$}+r_0%
\mbox{{\boldmath ${K}$}$_{n}$}/(r_1\phi_0)$, then we have
\begin{align}
&D[Q,P]=\frac{1}{2}\left\{-\log|\hat{\mbox{\boldmath ${K}$}}_{n}^{-1} %
\mbox{{\boldmath ${K}$}$_{n}$}|+\log|\mbox{\boldmath ${B}$}|+Tr(\hat{%
\mbox{\boldmath ${K}$}}_{n}^{-1} \mbox{{\boldmath ${K}$}$_{n}$}%
\mbox{{\boldmath ${B}$}$^{-1}$})\right.  \notag \\
&\hskip 2cm \left.+r_0||f_{0}||^2_k+r_0\mbox{\boldmath ${\alpha }$} %
\mbox{{\boldmath ${K}$}$_{n}$}(\hat{\mbox{\boldmath ${K}$}}_{n}^{-1} %
\mbox{{\boldmath ${K}$}$_{n}$}-\mbox{{\boldmath ${I}$}$_{n}$})%
\mbox{\boldmath ${\alpha}$}-n\right\},  \label{KBdist} \\
&E_{Q}(-\log p_{\phi_0,r}(\mbox{{\boldmath ${y}$}$$}|\tilde{f},%
\mbox{\boldmath ${X}$}))  \notag \\
\leq& -\log p_{\phi_0,r}(\mbox{{\boldmath ${y}$}$$}|f_0,%
\mbox{\boldmath
${X}$}))+\frac{r_0}{2r_1\phi_0}Tr( \mbox{{\boldmath ${K}$}$_{n}$}%
\mbox{{\boldmath ${B}$}$^{-1}$}),  \label{EQP}
\end{align}
where $\hat{\mbox{\boldmath ${K}$}}_{n}=(k(\mbox{{\boldmath ${x}$}$_{j}$},%
\mbox{{\boldmath ${x}$}$_{l}$};\hat{\mbox{\boldmath ${\theta}$}}))_{n\times
n}$.

From (\ref{FLD}), (\ref{KBdist}) and (\ref{EQP}), it gives that
\begin{align}  \label{GPcons-1}
&-\log p_{\phi_0,\theta, r}(\mbox{{\boldmath ${y}$}$$}|%
\mbox{{\boldmath
${X}$}$$})+\log p_{\phi_0,r}(\mbox{{\boldmath ${y}$}$$}|f_0,%
\mbox{\boldmath
${X}$}))  \notag \\
\leq& \frac{1}{2}\left\{-\log|\hat{\mbox{\boldmath ${K}$}}_{n}^{-1} %
\mbox{{\boldmath ${K}$}$_{n}$}|+\log|\mbox{\boldmath ${B}$}|+Tr((\hat{%
\mbox{\boldmath ${K}$}}_{n}^{-1} \mbox{{\boldmath ${K}$}$_{n}$}+r_0 %
\mbox{{\boldmath ${K}$}$_{n}$}/(r_1\phi_0))\mbox{{\boldmath ${B}$}$^{-1}$})
+r_0||f_{0}||^2_k\right.  \notag \\
&\left.+r_0\mbox{\boldmath ${\alpha }$} \mbox{{\boldmath ${K}$}$_{n}$}(\hat{%
\mbox{\boldmath ${K}$}}_{n}^{-1} \mbox{{\boldmath ${K}$}$_{n}$}-%
\mbox{{\boldmath ${I}$}$_{n}$})\mbox{\boldmath ${\alpha}$}-n\right\}.
\end{align}
Due to the bounded, continuous covariance function and almost sure
convergence of $\hat{\mbox{\boldmath ${\theta}$}}$, we have $\hat{%
\mbox{\boldmath ${K}$}}_{n}^{-1} \mbox{{\boldmath ${K}$}$_{n}$} -%
\mbox{{\boldmath ${I}$}$_{n}$}\rightarrow 0$ as $n\rightarrow \infty$.
Hence, there exist positive constants $c$ and $\varepsilon$ such that for a
large enough $n$
\begin{align}  \label{eqns}
&-\log|\hat{\mbox{\boldmath ${K}$}}_{n}^{-1} \mbox{{\boldmath ${K}$}$_{n}$}%
|<c,~~\mbox{\boldmath ${\alpha }$} \mbox{{\boldmath ${K}$}$_{n}$}(\hat{%
\mbox{\boldmath ${K}$}}_{n}^{-1} \mbox{{\boldmath ${K}$}$_{n}$}-%
\mbox{{\boldmath ${I}$}$_{n}$})\mbox{\boldmath ${\alpha}$}<c,  \notag \\
&Tr(\hat{\mbox{\boldmath ${K}$}}_{n}^{-1} \mbox{{\boldmath ${K}$}$_{n}$} %
\mbox{{\boldmath ${B}$}$^{-1}$})<Tr((\mbox{{\boldmath ${I}$}$_{n}$}%
+\varepsilon \mbox{{\boldmath ${K}$}$_{n}$})\mbox{{\boldmath ${B}$}$^{-1}$}).
\end{align}
Plugging (\ref{eqns}) in (\ref{GPcons-1}), we have the inequality (\ref{Aeq1}%
). $\sharp$

\vskip 10pt

Under condition \newline
(A) $||f_{0i}||_{k}$ is bounded and $E_{\mbox{{\boldmath ${X}$}$_{i}$}}(\log|%
\mbox{{\boldmath ${I}$}$_{n}$}+c \mbox{{\boldmath ${K}$}$_{in}$}|)=o(n)$ for
any $c>0$ and $i=1,...,I$,\newline
it follows from Lemma 1 that for $I=J=1$,
\begin{equation}
\frac{1}{n}E_{\mbox{{\boldmath ${X}$}$$}}(D[p_{\phi _{0}}(%
\mbox{{\boldmath
${y}$}$$}|f_{0},\mbox{{\boldmath ${X}$}$$}),p_{\phi _{0},\hat{\theta},\hat r
}(\mbox{{\boldmath ${y}$}$$}|\mbox{{\boldmath ${X}$}$$})])\longrightarrow 0,{%
\mbox{as}}~~n\rightarrow \infty.  \label{inforcon}
\end{equation}

\vskip10pt \noindent \textbf{Proof of Theorem 1}: Suppose $%
\mbox{{\boldmath
${y}$}$_{i}$}=\{y_{i1},...,y_{in}\}$ are generated from model (\ref{assumed}%
) with the mean function $h(\mbox{\boldmath ${x}$})=0$, and covariance
kernel $k_i$ is bounded and continuous in parameter $%
\mbox{{\boldmath
${\theta}$}$_{i}$}$. Under condition (A), similar to proof of (\ref{inforcon}%
), we show that
\begin{equation}
\frac{1}{n} E_{\mbox{{\boldmath ${X}$}$_{i}$}}(D[p_{\phi _{0i}}(%
\mbox{{\boldmath ${y}$}$_{i}$}|f_{0i},\mbox{{\boldmath ${X}$}$_{i}$}%
),p_{\phi _{0i},\hat{\theta}_i,\hat r_i}(\mbox{{\boldmath ${y}$}$_{i}$}|%
\mbox{{\boldmath ${X}$}$_{i}$})])\longrightarrow 0,{\mbox{as}}~~n\rightarrow
\infty.  \notag
\end{equation}
Hence, Theorem 1 holds.$\sharp$


\newpage

\begin{table}[!ht]
\caption{MSEs and standard deviations (in parentheses) of the predictions
from GPR, eTPR and GP-TP }
\label{tab1-1}\tabcolsep=4pt\fontsize{9}{16}\selectfont
 \vskip 0pt
\par
\begin{center}
\begin{tabular}{cccccc}
\hline
Error & Disturbance & $\gamma$ & GPR & eTPR & GP-TP \\ \hline
Gaussian & constant & 0.5 & 0.038(0.022) & 0.038(0.022) & 0.037(0.021) \\
&  & 1.0 & 0.060(0.027) & 0.060(0.028) & 0.038(0.021) \\
&  & 2.0 & 0.145(0.052) & 0.145(0.053) & 0.044(0.027) \\
Gaussian & random & 0.5 & 0.157(0.394) & 0.156(0.391) & 0.047(0.118) \\
&  & 1.0 & 0.177(0.536) & 0.177(0.536) & 0.044(0.029) \\
&  & 2.0 & 0.225(0.425) & 0.225(0.426) & 0.051(0.049) \\
ETP($\nu=2.0$) & constant & 0.5 & 0.049(0.033) & 0.049(0.032) & 0.040(0.028)
\\
&  & 1.0 & 0.072(0.044) & 0.072(0.044) & 0.047(0.033) \\
&  & 2.0 & 0.150(0.066) & 0.151(0.066) & 0.050(0.035) \\
ETP($\nu=1.0$) & constant & 0.5 & 0.089(0.155) & 0.089(0.157) & 0.048(0.034)
\\
&  & 1.0 & 0.093(0.081) & 0.093(0.082) & 0.055(0.043) \\
&  & 2.0 & 0.167(0.126) & 0.167(0.126) & 0.058(0.045) \\ \hline
\end{tabular}
\end{center}
\end{table}

\begin{table}[!ht]
\caption{MSEs and standard deviations (in parentheses) of the predictions
from GPR and GP-TP, where the error term has Gaussian distribution}
\label{tab1}\tabcolsep=4pt\fontsize{9}{16}\selectfont
 \vskip 0pt
\par
\begin{center}
\begin{tabular}{cccc}
\hline
Disturbance & $\gamma$ & GPR & GP-TP \\ \hline
constant & 0.5 & 0.037(0.015) & 0.036(0.015) \\
& 1.0 & 0.061(0.020) & 0.040(0.017) \\
& 2.0 & 0.144(0.037) & 0.051(0.026) \\
random & 0.5 & 0.276(1.212) & 0.045(0.042) \\
& 1.0 & 0.182(0.407) & 0.048(0.049) \\
& 2.0 & 4.364(92.235) & 0.052(0.036) \\ \hline
\end{tabular}
\end{center}
\end{table}

\begin{table}[!ht]
\caption{Estimation of random effects $r_{ij}$ and their standard deviations
(in parentheses) from GP-TP for $i=1,2$ and $j=1,..,6$.}
\label{tab2}\tabcolsep=3pt\fontsize{9}{12}\selectfont
 \vskip 0pt
\par
\begin{center}
\begin{tabular}{cccccccc}
\hline
Disturbance & $\gamma$ & $r_{11}$ & $r_{12}$ & $r_{13}$ & $r_{14}$ & $r_{15}$
& $r_{16}$ \\ \hline
constant & 0.5 & 0.534(0.284) & 0.531(0.284) & 0.536(0.285) & 0.554(0.295) &
0.539(0.288) & 0.819(0.433) \\
& 1.0 & 0.313(0.278) & 0.318(0.280) & 0.315(0.279) & 0.311(0.273) &
0.319(0.289) & 1.113(0.969) \\
& 2.0 & 0.233(0.209) & 0.235(0.213) & 0.232(0.208) & 0.23(0.204) &
0.234(0.206) & 2.553(2.107) \\
random & 0.5 & 0.357(0.292) & 0.367(0.295) & 0.349(0.284) & 0.353(0.285) &
0.356(0.289) & 3.707(20.884) \\
& 1.0 & 0.313(0.291) & 0.310(0.284) & 0.308(0.289) & 0.318(0.298) &
0.303(0.284) & 2.187(5.027) \\
& 2.0 & 0.254(0.244) & 0.258(0.252) & 0.255(0.251) & 0.254(0.241) &
0.255(0.242) & 4.019(17.145) \\
&  & $r_{21}$ & $r_{22}$ & $r_{23}$ & $r_{24}$ & $r_{25}$ & $r_{26}$ \\
\cline{3-8}
constant & 0.5 & 0.549(0.29) & 0.548(0.297) & 0.525(0.276) & 0.545(0.3) &
0.537(0.292) & 0.855(0.456) \\
& 1.0 & 0.315(0.281) & 0.319(0.286) & 0.319(0.285) & 0.328(0.296) &
0.322(0.289) & 1.121(0.975) \\
& 2.0 & 0.236(0.208) & 0.229(0.204) & 0.238(0.21) & 0.229(0.201) &
0.227(0.202) & 2.618(2.145) \\
random & 0.5 & 0.369(0.313) & 0.371(0.299) & 0.362(0.302) & 0.362(0.293) &
0.364(0.298) & 3.447(30.013) \\
& 1.0 & 0.317(0.297) & 0.316(0.301) & 0.304(0.279) & 0.311(0.298) &
0.305(0.279) & 2.753(11.809) \\
& 2.0 & 0.251(0.241) & 0.25(0.241) & 0.243(0.231) & 0.25(0.24) & 0.25(0.245)
& 4.193(15.459) \\ \hline
\end{tabular}
\end{center}
\end{table}

\begin{table}[!ht]
\caption{Estimation of random effects from GP-TP for the game data}
\label{tab4}\tabcolsep=3pt\fontsize{9}{12}\selectfont
 \vskip 0pt
\par
\begin{center}
\begin{tabular}{ccccccccccccccc}
\hline
group & instruction & r1 & r2 & r3 & r4 & r5 & r6 & r7 & r8 & r9 & r10 & r11
& r12 & r13 \\ \hline
Young & single & 0.183 & 0.569 & 0.264 & 0.379 & 0.369 & 0.460 & 1.016 &
0.281 & 0.367 & 0.206 & 0.280 & 0.450 &  \\
& double & 0.161 & 0.787 & 0.366 & 9.472 & 0.796 & 0.310 & 1.373 & 0.259 &
0.147 & 0.101 & 0.133 & 1.051 &  \\
& double-4 & 0.309 & 1.529 & 0.710 & - & 1.546 & 0.609 & 2.655 & 0.507 &
0.274 & 0.195 & 0.251 & 2.040 &  \\ \cline{2-15}
Old & single & 0.092 & 0.241 & 0.767 & 0.238 & 0.063 & 0.719 & 0.288 & 0.143
& 0.172 & 0.294 & 1.115 & 0.952 & 5.193 \\
& double & 0.268 & 0.205 & 1.335 & 2.192 & 0.450 & 0.135 & 0.213 & 1.303 &
0.131 & 0.146 & 0.227 & 0.147 & 0.371 \\
& single-13 & 0.169 & 0.463 & 1.488 & 0.464 & 0.116 & 1.410 & 0.546 & 0.263
& 0.319 & 0.561 & 2.159 & 1.868 & - \\ \hline
\end{tabular}
\end{center}
\end{table}

\end{document}